\documentclass[reprint,showkeys,aip,jap]{revtex4-1}

\usepackage[pdftex]{graphicx}% Include figure files
\usepackage{dcolumn}% Align table columns on decimal point
\usepackage{bm}% bold math

\usepackage{color}

\begin{document}
\title{Electrochemical membrane microactuator with a millisecond response time }

\author{Ilia V. Uvarov}
\affiliation{Yaroslavl Branch of the Institute of Physics and Technology RAS, 150007, Universitetskaya 21, Yaroslavl, Russia}

\author{Mikhail V. Lokhanin}
\affiliation{P. G. Demidov Yaroslavl State University, Sovetskaya 14, 150000 Yaroslavl, Russia}

\author{Alexander V. Postnikov}
\affiliation{Yaroslavl Branch of the Institute of Physics and Technology RAS, 150007, Universitetskaya 21, Yaroslavl, Russia}

\author{Artem E. Melenev}
\affiliation{P. G. Demidov Yaroslavl State University, Sovetskaya 14, 150000 Yaroslavl, Russia}

\author{Vitaly B. Svetovoy}
\email[Corresponding author: ]{v.b.svetovoy@rug.nl}
\affiliation{Zernike Institute for Advanced Materials, University of Groningen - Nijenborgh 4, 9747 AG Groningen, The Netherlands}
\affiliation{Yaroslavl Branch of the Institute of Physics and Technology RAS, 150007, Universitetskaya 21, Yaroslavl, Russia}

%\date{\today}

\begin{abstract}
Lack of fast and strong actuators to drive microsystems is well recognized. Electrochemical actuators are considered attractive for many applications but they have long response time (minutes) due to slow gas termination. Here an electrochemical actuator is presented for which the response time can be as short as $1\:$ms. The alternating polarity water electrolysis is used to drive the device. In this process only nanobubbles are formed. The gas in nanobubbles can be terminated fast due to surface assisted reaction between hydrogen and oxygen that happens at room temperature. The working chamber of the actuator contains concentric titanium electrodes; it has a diameter of $500\:\mu$m and a height of $8\:\mu$m. The chamber is sealed by a polydimethylsiloxane (PDMS) membrane of $30\:\mu$m thick. The device is characterized by an interferometer and a fast camera. Cyclic operation at frequency up to $667\:$Hz with a stroke of about 30\% of the chamber volume is demonstrated. The cycles repeat themselves with high precision providing the volume strokes in picoliter range. Controlled explosions in the chamber can push the membrane up to $90\:\mu$m.
\end{abstract}


\maketitle

\section{Introduction}

Miniaturization of different systems is a clear trend in the modern world. To drive any mechanical device one has to transform the driving energy (for example, energy of a battery) to mechanical movement. This role is played by actuators. The actuator is only a part of microsystems but it is a bulky part dissipating a large portion of the driving energy. In contrast with the macroscopic world, where we have got efficient and reliable motors such as electromagnetic motors or internal combustion engines, there is obvious lack of fast and strong autonomous microactuators. For example, microfluidic devices \cite{Whitesides2006} that are designed for fast and cheap chemical, biological, or medical tests often rely on macroscopic compressors to pump liquid through microchannels. Fabrication of a small field coil to generate a significant magnetic field in the electromagnetic motors is not an easy task \cite{Buttgenbach2014}. On the other hand, microengines using internal combustion \cite{Maruta2011} cannot be scaled down because combustion reactions quench in a small volume due to fast heat escape via the volume boundaries \cite{Veser2001,Fernandez2002}.

A great number of principles and techniques is used to build microactuators \cite{Abhari2012,Ashraf2011,Au2011}. They include piezoelectric \cite{Lee2004,Sayar2012}, electrostatic \cite{Lee2012,Conrad2015}, thermal \cite{Sim2003,Chia2011,Chan2010}, electrokinetic \cite{Chen2007,Chen2008} and many other actuation principles.  Actuators using electrostatic forces are fast but they develop rather weak forces. On the contrary, the actuators using the thermal principle are slow but strong. Piezoelectric actuators are very popular, they are strong and fast but the main disadvantage is a small stroke per unit voltage. Due to this feature one has to apply a high voltage or has to increase the size of the actuator. For example, in a recent commercial product Insulin Nanopump$^{\textrm{TM}}$ developed by Debiotech a piezo actuator with a diameter of $5\:$mm is used, which is by far the largest element of the pump \cite{Piveteau2013}. An additional problem is that the piezoelements are not well compatible with standard microfabrication technology.

A special class of actuators uses the electrochemical process such as water decomposition to produce mechanical work \cite{Neagu1996,Cameron2002,Hua2002,Ateya2004,Meng2008,Kjeang2009,Li2010,Yi2015}. In these devices the gas generated by water electrolysis pushes a flexible membrane. The electrochemical actuation develops large driving force, provides accurate flow control, generates low heat, and is compatible with microtechnology. However, the electrochemical actuation is notoriously slow because to return the actuator in the initial state one has to terminate all the produced gas. It can be done by natural dissolution of the gas in a large volume of liquid that takes a long time ($5-10\:$ minutes) \cite{Neagu1996,Cameron2002}, some authors using venting of the bubbles \cite{Meng2008}, but the most advanced way is to initiate the reverse reaction between the gases, for example, using catalytic properties of platinized electrodes \cite{Yi2015}. In the latter case the response time is considerably reduced but stays in the range of $100\:$s.

Alternating polarity electrolysis performed by short (a few $\mu$s) voltage pulses \cite{Svetovoy2011} demonstrated unexpected properties that can be used as a new principle to build a fast electrochemical actuator \cite{Svetovoy2014}. If each electrode is kept equal time at positive and at negative potential and if the polarity changes with a frequency of the order of $100\:$kHz, only nanobubbles are generated by the process. Production of a large amount of gas was proven by the change of the refractive index of liquid around the electrodes \cite{Svetovoy2011}. This amount can be so large that the optical image of the electrodes is distorted \cite{Postnikov2017} but no light scattering on microscopic or larger bubbles is observed. The alternating polarity process builds up a significant pressure in a closed microchamber \cite{Svetovoy2014} but, again, no light scattering on microscopic bubbles were observed. The most peculiar feature of the process is that this pressure relaxes just in $100\:\mu$s after switching off the electrical pulses.

Fast termination of the gas is explained by combustion of hydrogen and oxygen in nanobubbles \cite{Svetovoy2011}. The reaction happens spontaneously at room temperature due to surface assisted processes that dominate in nanobubbles due to a large surface-to-volume ratio \cite{Prokaznikov2017}. Although the details of the reaction mechanism are still not clear, one can use this reaction to reduce drastically the response time of the electrochemical actuators. In this paper we are going to show that this time can be reduced from minutes to $1\:$ms and, probably, this is not the ultimate limit.

The first practical application of the alternating polarity electrolysis for a valveless micropump was not very successful \cite{Uvarov2016a}. The pump demonstrated a low pumping rate of $0.15\:$nl/min at an operating frequency of $4\:$Hz. The reaction chamber of the device was made from polydimethylsiloxane (PDMS) that is too soft material: instead of pumping liquid the chamber is inflated in response to the pressure increase. Moreover, platinum electrodes were not able to support a high current density and destroyed fast.

A miropump of the same design works much better in a so called regime of exploding bubbles \cite{Uvarov2017}. In this case the pumping rate was $12\:$nl/min at an operating frequency of $1\:$Hz that is not bad keeping in mind that the chamber volume ($3.7\:$nl) is much smaller than in most of the micropumps. Fast video reviled the processes happening in the chamber during the pumping. For each cycle there is an incubation time needed to reach a critical density of nanobubbles inside of the chamber. When this critical concentration is reached the nanobubbles start to merge and form a microbubble containing stoichiometric mixture of H$_2$ and O$_2$. This mixture is ignited spontaneously and the bubble explodes. As the result just in $100\:\mu$s all liquid is pushed out of the chamber providing the pumping. Titanium electrodes used in \cite{Uvarov2017} demonstrated much better stability at high current density than Pt electrodes.

Here we describe a membrane actuator that uses as the driving process the alternating polarity electrolysis. The device is tested in different regimes.

\section{Experimental}

\begin{figure}[tbhp]
\centering
\includegraphics[width=.99\linewidth]{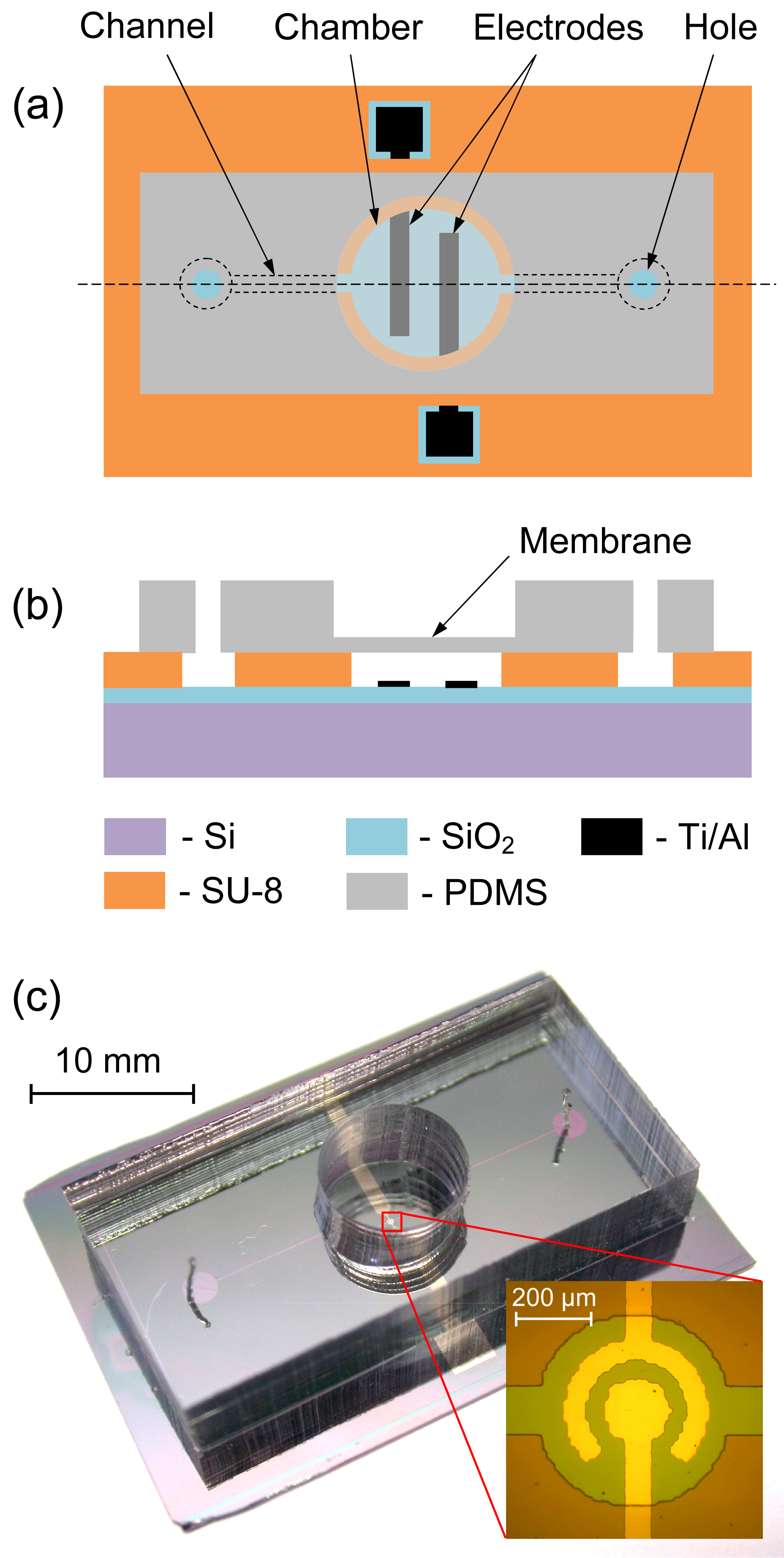}
\caption{Design of the actuator: (a) the top view; (b) cross section along the dashed horizontal line in (a). Ready to use actuator is presented in panel (c). The inset shows a zoomed view of the working chamber.}
\label{fig:actuator}
\end{figure}

Design of the actuator and a ready-to-use device are shown in Fig.$\:$\ref{fig:actuator}. The device has a working chamber with metallic electrodes inside, channels to fill the system with the electrolyte, and holes for connection with microfluidic tubes. The diameter of the chamber is $500\:\mu$m and its height is $8\:\mu$m. An oxidized silicon substrate is used as the bottom wall, the side walls are made of cured SU-8 resist, and the upper wall is a flexible PDMS membrane with a thickness of $30\:\mu$m. The electrodes have a circular shape; the external electrode has a larger diameter of $380\:\mu$m and the internal one is $180\:\mu$m in diameter. The electrodes contain two metallic layers deposited by magnetron sputtering. The bottom layer is $500\:$nm thick Al, which is necessary to reduce the resistance of the contact lines. As the working layer we are using titanium ($500\:$nm), which demonstrates very good stability at high current density flowing through the electrolyte. A molar solution of Na$_2$SO$_4$ in distilled water is used as the electrolyte. The actuator with a close-up view of the chamber is shown in Fig.$\:$\ref{fig:actuator}(c). The circular electrodes provide more homogeneous current density distribution in comparison with the interdigitated fingers used previously \cite{Uvarov2017}.

\begin{figure}[tbhp]
\centering
\includegraphics[width=.99\linewidth]{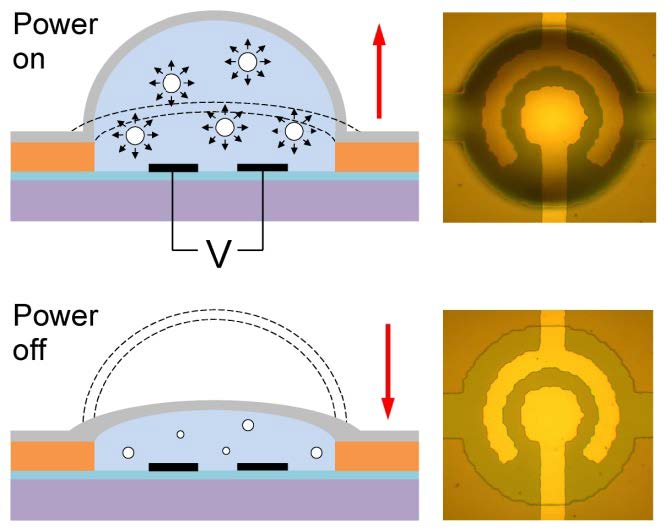}
\caption{Operation principle of the actuator. When the voltage pulses are switched on the gas produced in the chamber electrochemically pushes the membrane up. When the voltage is switched off the gas in nanobubbles is terminated fast and the membrane is going down. The images on the right show the chamber under extra pressure (top) and without it (bottom).}
\label{fig:principle}
\end{figure}

The actuator operates in the following way. A series of short ($\sim 1\:\mu$s) voltage pulses of alternating polarity is applied to the working electrode while the other one is grounded. These pulses produce hydrogen
and oxygen gases confined in nanobubbles, the pressure in the chamber increases and pushes the membrane up (Fig.$\:$\ref{fig:principle}). When the pulses are switched off, the gas in nanobubbles recombines fast back to water releasing the pressure and the membrane returns to its initial state. A short recombination time allows a significant reduction of the actuation cycle.

The device is fabricated in three stages. At the first stage a silicon substrate containing the electrodes, the chamber and channels is prepared. The silicon wafer is oxidized to grow the SiO$_2$ dielectric layer up to $0.9\:\mu$m thick. A $10\:$nm adhesive Ti layer, $500\:$nm Al layer, and $500\:$nm Ti layer are deposited by magnetron sputtering followed by the
UV photolithography and by the wet etching of metals. Thus, the electrodes are fabricated on the dielectric layer. Further, the walls of the chamber and channels are formed in $8\:\mu$m thick layer of SU-8 3005 photoresist. At the second stage a $4\:$mm thick PDMS (Sylgard 184, 10:1 mixture) structure holding a $30\:\mu$m thick membrane is fabricated on a separate
wafer. At the final stage the PDMS structure and the silicon substrate are assembled together using irreversible bonding of PDMS to SU-8 \cite{Zhang2011}.

Performance of the actuator is measured using a homemade homodyne Michelson interferometer with quadrature signals \cite{Reibold1981,Pozar2011}. The optical scheme is shown in Fig.$\:$1S (see Supplementary material). Laser beam (wavelength $\lambda = 0.63\:\mu$m) passed through a polarizing filter and a beamsplitter is focused on the center of the membrane by $10\times$ objective lens. To make the membrane reflective a $20\:$nm thick Al layer is deposited onto its surface. Photodetectors of the interferometer are based on photodiodes with transimpedance amplifiers. The bandwidth of the photodetectors is $1\:$MHz. The interferometer produces two phase-shifted output signals, which are computer processed to extract the membrane deflection.  An example of the raw signals of the interferometer and the restored deflection of the membrane is given in Supplementary material together with the equations necessary for the signal restorations. Deflection of the membrane is independently determined from the fast camera images.

The driving voltage is provided by a waveform generator Tektronix AFG1022 and amplified ten times using a homemade amplifier \cite{Uvarov2017}. The voltage, current flowing through the electrolyte, and the output signals
of the interferometer are recorded by a PicoScope 5000. In order to observe the processes in the chamber and the deflection of the membrane we are using a high speed camera Photron FASTCAM 1024PCI with a nominal frame rate of $10\:000\:$fps. The camera is triggered synchronously with the driving pulses.

\begin{figure}[tbhp]
\centering
\includegraphics[width=.99\linewidth]{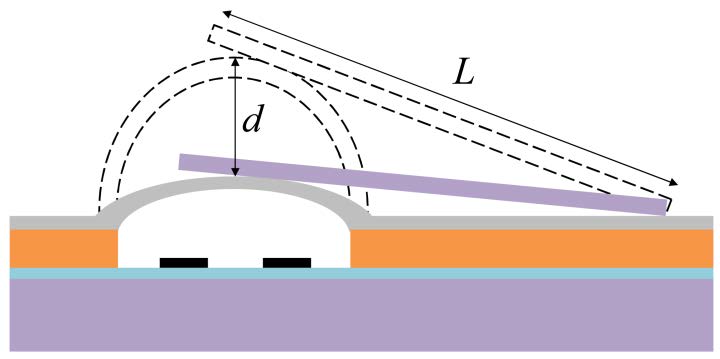}
\caption{Application of load to the membrane. The weight of a silicon mirror provides the load and the mirror surface is used to read out the displacement signal with the interferometer.}
\label{fig:load_scheme}
\end{figure}

To see how the actuator works under a load we put a small silicon mirror on the chamber as shown in Fig.$\:$\ref{fig:load_scheme}. The mirror has dimensions of $2000\times 840\times 30\:\mu$m$^3$. With the interferometer one can follow the movement of the mirror and find the deflection of the membrane under the load. The load force acting on the membrane from the mirror is estimated as $F=mg/2\approx 0.6\:\mu$N.

\section{Results and discussion}

\subsection{Actuation by a single series of pulses}
There are two possible responses of the actuator on the continuous application of the alternating polarity pulses to the electrodes. While the voltage amplitude $U$ is smaller than a threshold value $U_{th}$ the membrane reaches a maximum deflection and does not change anymore. In this steady state the gas formed by the Faraday current is balanced by the gas that gets into the recombination reaction \cite{Svetovoy2014}. When $U>U_{th}$ the concentration of nanobubbles in the chamber reaches a critical value and many of them merge to form a microbubble containing 2:1 mixture of H$_2$ and O$_2$ gases. This mixture explodes giving a strong push to the membrane. The gas in the chamber behaves similar to that observed during operation of the micropump \cite{Uvarov2017}.

\begin{figure}[tbhp]
\centering
\includegraphics[width=.99\linewidth]{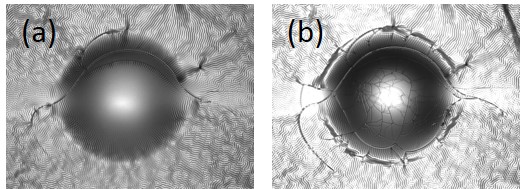}\\%d_vs_time
\includegraphics[width=.99\linewidth]{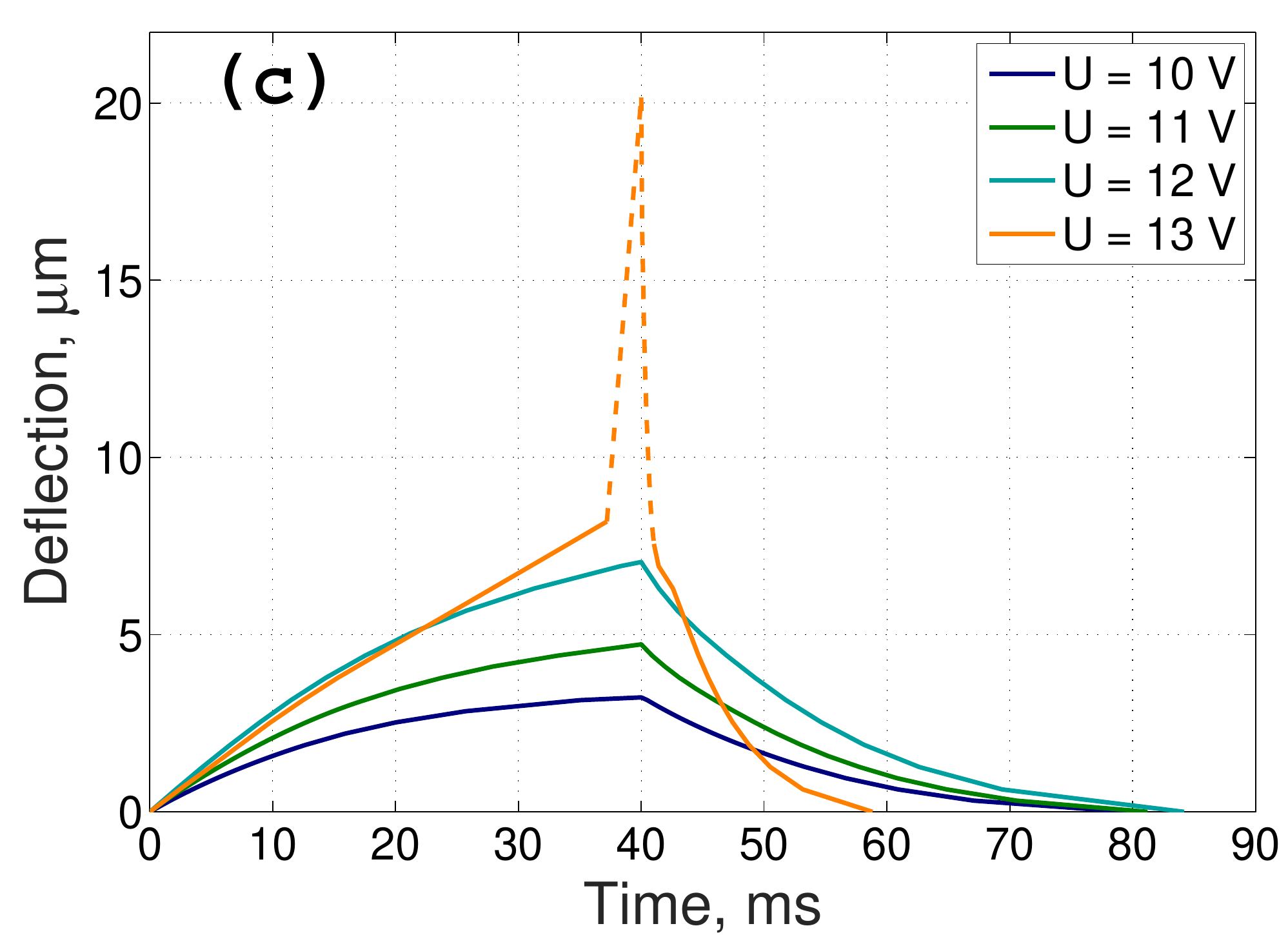}
\caption{(a) PDMS membrane after deposition of Al layer. (b) Cracks in Al layer on top of the membrane after an explosion in the chamber. (c) Deflection of the membrane in the center as a function of time for different voltage amplitudes at the driving frequency $f=500\:$kHz. The actuator is driven by a series of pulses of $40\:$ms long. For $U=13\:$V the explosion in the chamber strongly pushes the membrane up. For the dashed part of the curve the signals are changed too fast to be correctly recorded by the photodetectors.}
\label{fig:d_vs_time}
\end{figure}

Deflection of the membrane for a long series of $N=40\:000$ pulses as a function of time is shown in Fig.$\:$\ref{fig:d_vs_time}(c) for different voltage amplitudes. The deflection was measured by the interferometer after deposition of an opaque aluminum layer ($20\:$nm) on top of the membrane. Originally the metallic layer contains a few or no cracks (see Fig.$\:$\ref{fig:d_vs_time}(a)). Actuation without explosions in the chamber does not add new cracks but after an explosion many cracks appear on the membrane as shown in Fig.$\:$\ref{fig:d_vs_time}(b). The cracks arise in the metallic layer due to significant deflection of the PDMS membrane.

Each voltage period consists of one positive and one negative pulses, which are repeated with a driving frequency $f$. The series of $N$ pulses has the duration $t_a=N/2f$ (active time).  As one can see $t_a=40\:$ms is a sufficiently long time for which the membrane deflection nearly reaches the asymptotic value. This asymptotic deflection increases with the voltage amplitude and when the amplitude reaches the threshold value $U_{th}=13\:$V the membrane deflection jumps up in response to the explosion in the chamber. When the pulses are switched off the membrane returns back in $20-30\:$ms. It is interesting to note that in the case of explosion it returns to the initial state even faster in $10\:$ms or so.

Observing the membrane from the side one can see how it deflects in the vertical direction. Video S1 recorded at $6\:000\:$fps shows the side view of the membrane for $U=14\:$V, $f=500\:$kHz, and for the number of pulses in the series $N=20\:000$. The motion is slowed down 80 times. The maximum deflection measured from the video frames is estimated as $10.8\pm 0.7\:\mu$m. It is larger than in Fig.$\:$\ref{fig:d_vs_time}(c) due to higher voltage amplitude; the series is too short to produce an explosion. Video S2 was recorded at $3\:000\:$fps and is slowed down 50 times. It shows the transparent membrane from the top for the amplitude $U=12\:$V. One can see that during the process no bubbles strongly scattering light are formed but the deflection of the membrane is visible. According to Fig.$\:$\ref{fig:d_vs_time}(c) the maximum deflection in this case ($t_a=20\:$ms) is expected to be $5\:\mu$m.

\begin{figure}[tbhp]
\centering
\includegraphics[width=.99\linewidth]{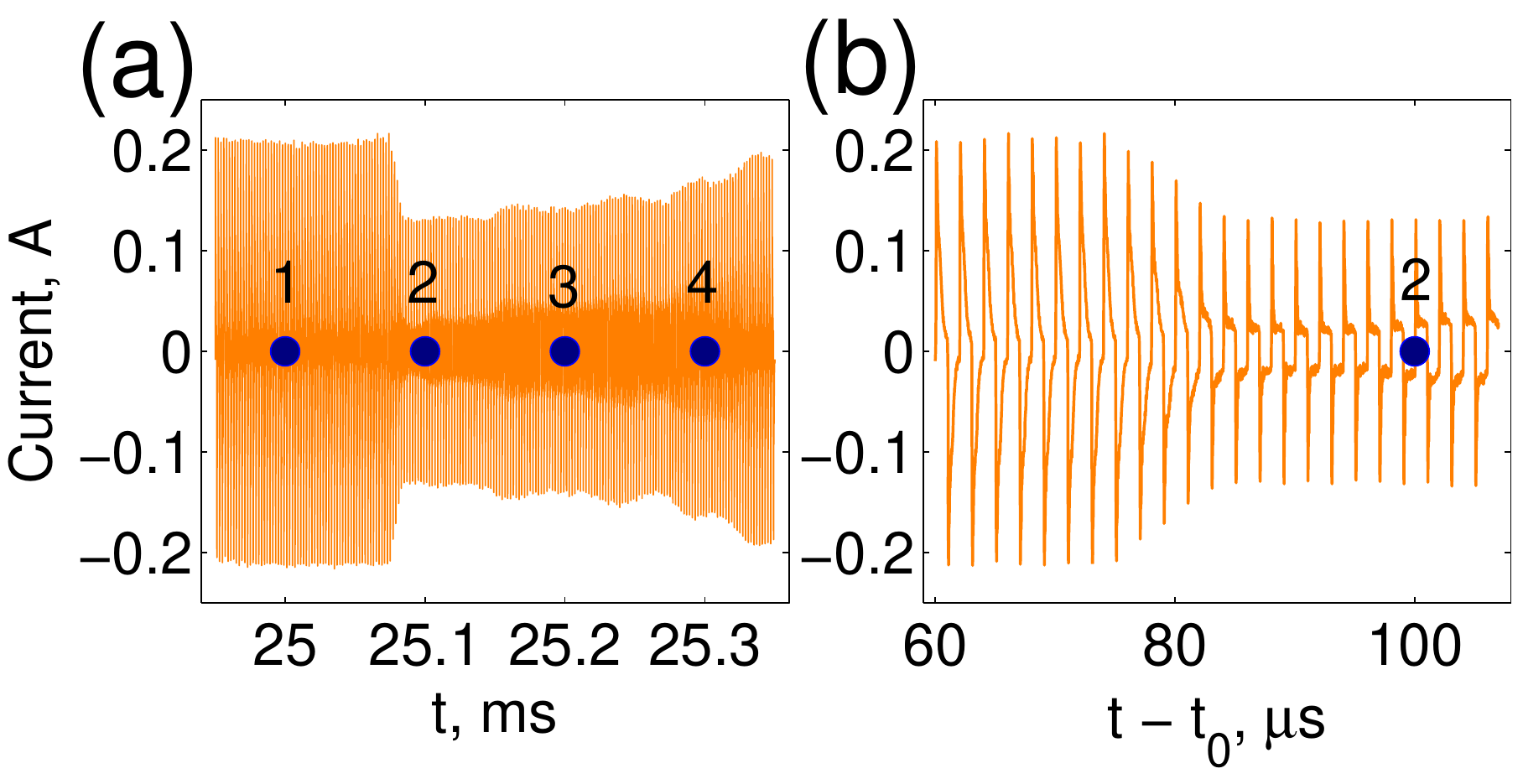}\\
\vspace{-0.1cm}
\includegraphics[width=.99\linewidth]{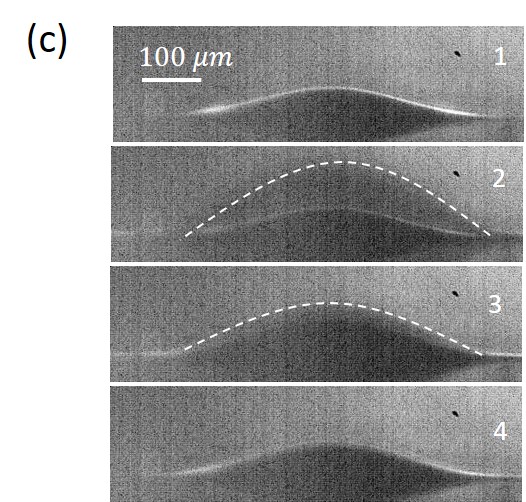}
\caption{(a) Current through the devise near the explosion. The parameters of the driving process are $U=15\:$V, $f=500\:$kHz, $N=26\:000$. The moment of explosion is indicated by the arrow; blue circles correspond to the positions of successive frames from the fast video. (b) Time-resolved current near the explosion; the moment $t_0$ corresponds to $25\:$ms. (c) Successive frames from Video S3 around the explosion show the side view of the membrane; maximum deflection one can see in frame 2, it is not sharp due to motion blur. For better visibility the maximum deflection in frames 2 and 3 is shown by white dashed lines. }
\label{fig:explosion}
\end{figure}

The interferometer is not fast enough to follow the deflection of the membrane during the explosion. For this reason a part of the deflection curve at $U=13\:$V is shown by the dashed line. The current behavior near the explosion (for a different sample) is demonstrated in Fig.$\:$\ref{fig:explosion}. A bubble growing in the chamber \cite{Uvarov2017} results in the current drop that occurs for $10\:\mu$s. During this time the bubble becomes larger than the diameter of the external electrode ($380\:\mu$m) in the lateral direction and blocks the current. In the vertical direction it grows much larger than the height of the chamber. In $1\:\mu$s ($1\:$MHz bandwidth) the membrane moves the distance more than $\lambda/4$ and the interferometer is not able to reproduce the movement correctly. The fast camera is also not able to follow the deflection but it gives more information on the amplitude of the deflection due to visible motion blur. A few successive frames made in the moments close to the explosion are shown in Fig.$\:$\ref{fig:explosion}. The maximum deflection of the membrane in frame 2 in comparison with frame 1 is estimated as $90\:\mu$m. All the frames one can see in Video S3 that was recorded at $10\:000\:$fps and is slowed down 667 times.

\begin{figure}[tbhp]
\centering
\includegraphics[width=.99\linewidth]{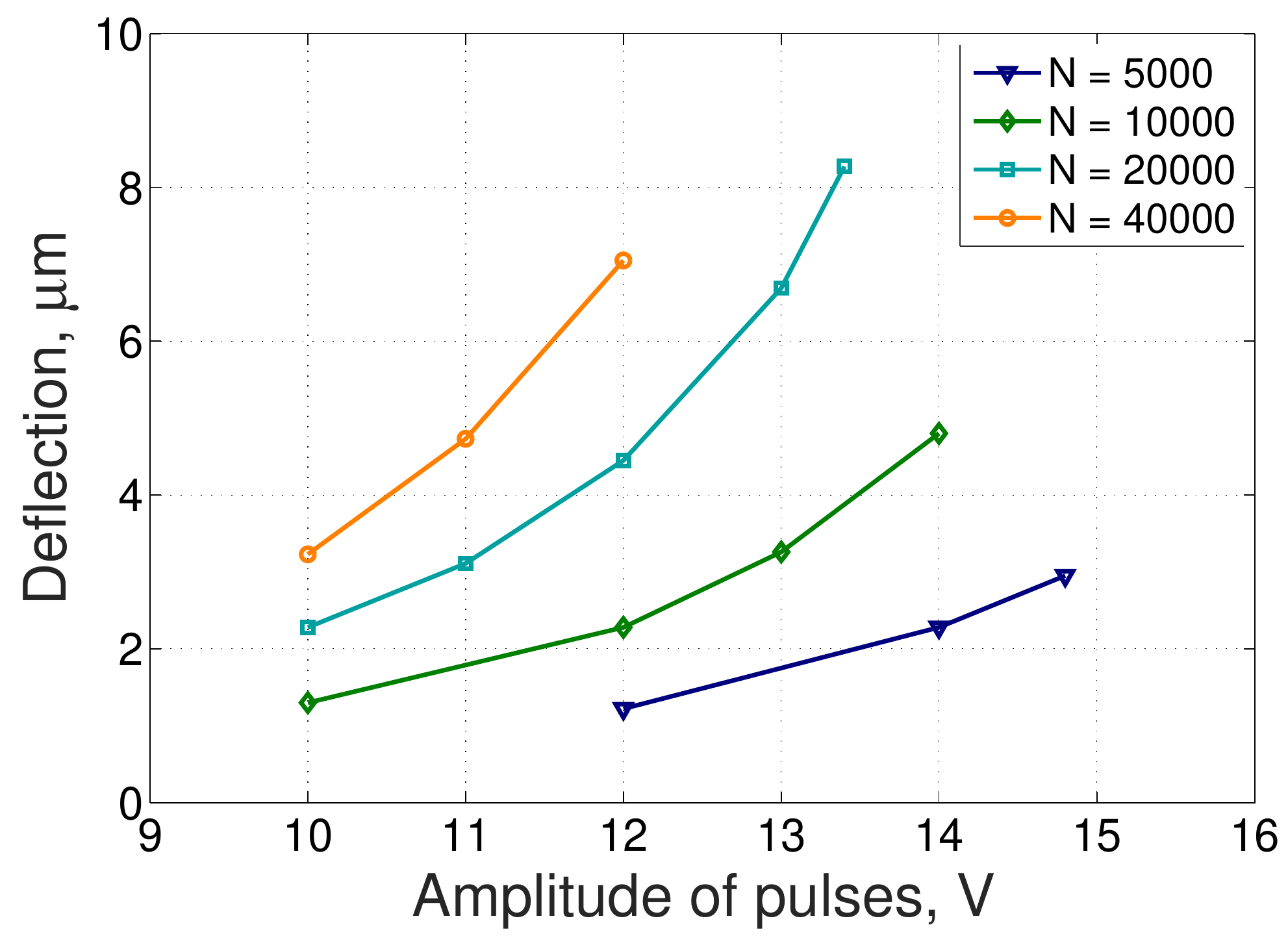}%Samples
\caption{Maximum deflection of the membrane as a function of voltage amplitude for different number of pulses $N$ in the series at the driving frequency $f=500\:$kHz.}
\label{fig:d_vs_U}
\end{figure}

The explosion can happen only after an incubation time $\tau$ needed to reach the critical concentration of nanobubbles. If the series of $N$ pulses is shorter than the incubation time $t_a<\tau$, then the maximum deflection of the membrane monotonously increases with $N$. It increases also with the voltage amplitude as shown in Fig.$\:$\ref{fig:d_vs_U}. One can see that the deflections in the range $2-7\:\mu$m are easily accessible. The threshold voltage is usually in the range $U_{th}=11-13\:$V but for some samples it can be smaller or larger.

The possibility to use explosions in the chamber for the cyclic operation of the actuator is very attractive. Explosions provide large strokes and can be repeated with a high frequency due to fast termination of the gas. However, in this paper we do not analyse such a regime due to the following reason. For realization of periodic explosions one has to choose the length of the series in such a way that the explosion happens at the very end of the series. Otherwise there are aftershocks produced by smaller exploding bubbles and these aftershocks happen randomly. In Video S3 the main explosion happens close to the end of the series but still one can see the aftershocks. Moreover, these smaller uncorrelated explosions sooner or later result in the formation of pinned bubbles, which do not disappear but only grow with time. Both of the problems can be resolved with a feedback control of the pulses but the results will be presented elsewhere.

\subsection{Cyclic actuation below the explosion threshold}

\begin{figure}[tbhp]
\centering
\includegraphics[width=.99\linewidth]{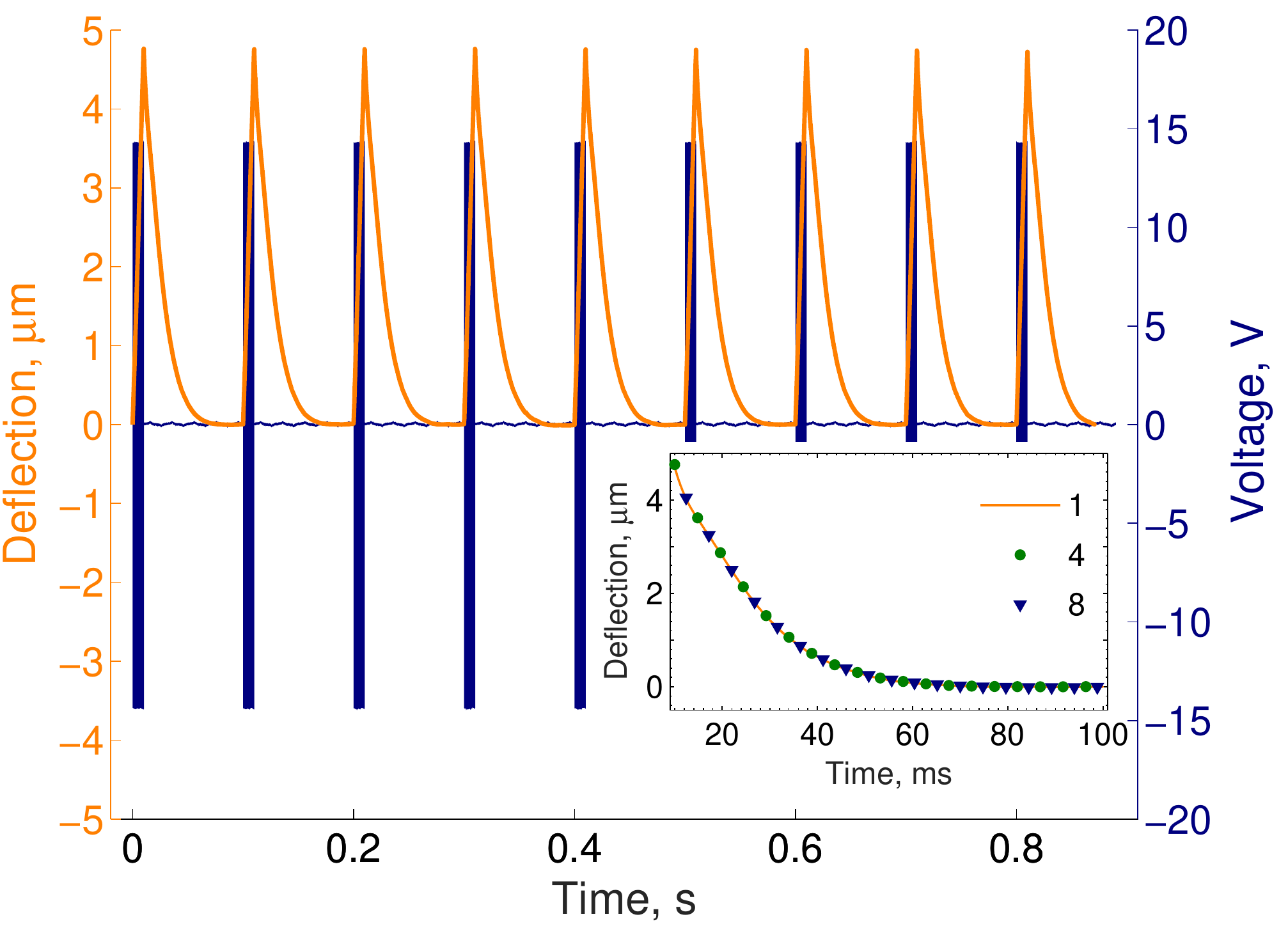}%fc_10Hz
\caption{Cyclic operation of the actuator at $f_c=10\:$Hz. Trains of voltage pulses are shown blue (right axis). The corresponding membrane deflection is shown light brown (left axis). The inset demonstrates the details of the relaxation curve for the cycles 1, 4, and 8 shifted to the same moment of time $t=10\:$ms.}
\label{fig:fc_10Hz}
\end{figure}

In this paper we concentrate on cyclic actuation below the explosion threshold. The actuators were tested in the cyclic operation mode when series of $N$ pulses were repeated with the operating frequency $f_c$. For a relatively low frequency $f_c=10\:$Hz the driving pulses and corresponding membrane deflection are shown in Fig.$\:$\ref{fig:fc_10Hz}. The process is driven at $U=14\:$V and $f=500\:$kHz by a series of $N=10\:000$ pulses. During the active time $t_a=10\:$ms the membrane deflects $4.8\:\mu$m that is more than half of the chamber height, then the pulses are switched off during $t_p=90\:$ms (passive time). The membrane deflection relaxes by half in $18\:$ms and returns to the initial state in $50\:$ms as one can see in the inset. An important feature of the process is a high stability of the operation: the deflection nearly exactly repeats itself in every cycle on the investigated time scale. A few very first cycles can be different from the others because the chamber has to be saturated with the nanobubbles.

Assuming parabolic deflection of the membrane the volume increase of the chamber in one cycle is $\Delta V = \pi R^2 d_0/2 \approx 0.47\:$nl, where $R=250\:\mu$m is the chamber radius and $d_0 = 4.8\:\mu$m is the stroke of the membrane. The volume of the chamber increases to one-third (the chamber volume is $1.57\:$nl). If this actuator would drive a micropump, the flow rate would be $0.28\:\mu$l/min in spite of a tiny volume of the chamber. The dose of the liquid delivered with every cycle is $0.47\:$nl/cycle and, probably, it can be made much smaller. These parameters are attractive for drug delivery systems.

\begin{figure}[tbhp]
\centering
\includegraphics[width=.99\linewidth]{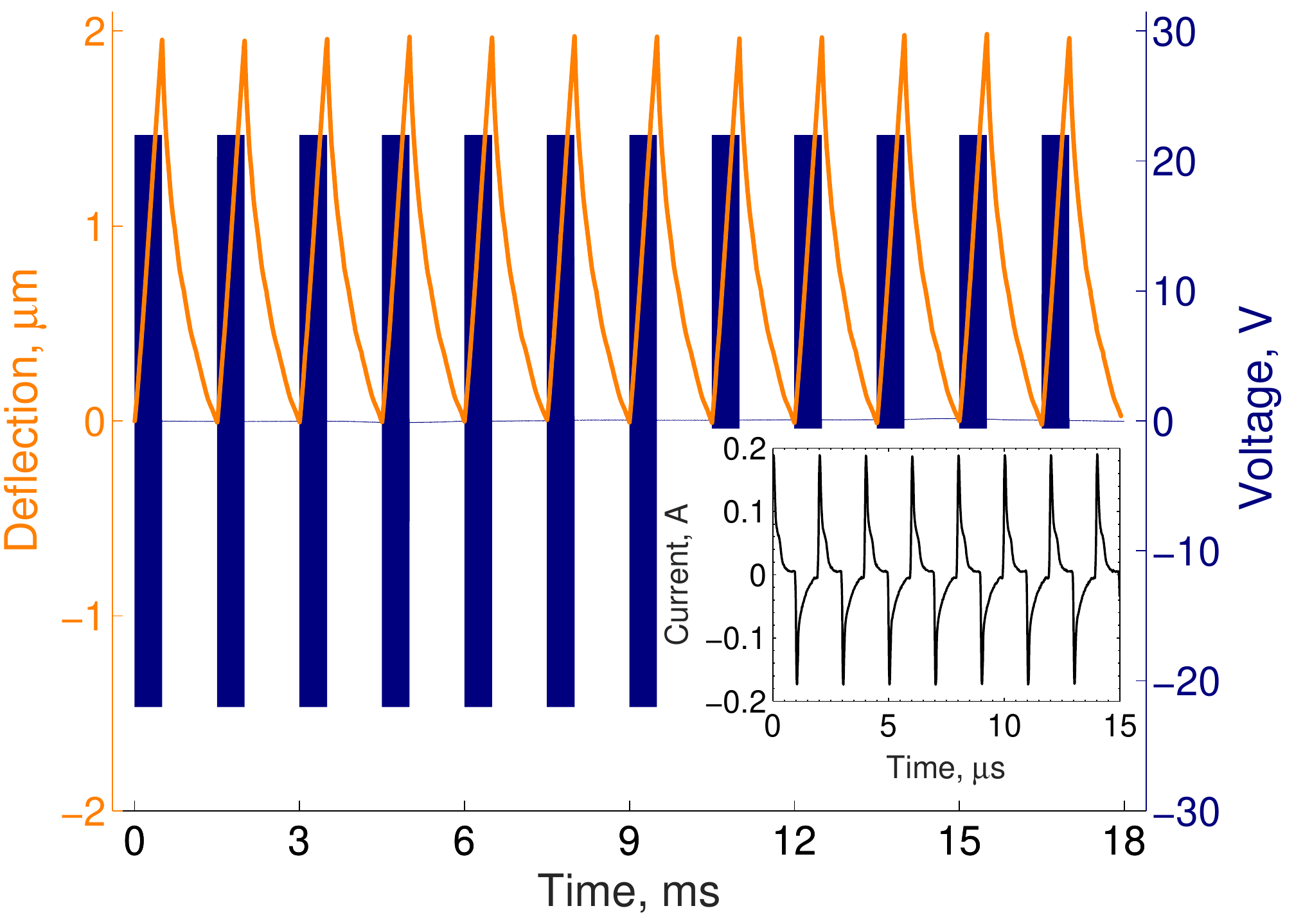}%fc_667Hz
\caption{Fast cyclic operation of the actuator at $f_c=667\:$Hz. Voltage pulses are shown blue (right axis). The corresponding membrane deflection is shown light brown (left axis). The inset shows the time-resolved current through the electrolyte.}
\label{fig:fc_667Hz}
\end{figure}

The cycle frequency can be significantly increased further but the price for this is increase in the voltage amplitude and some reduction of the stroke. An example of actuation at $f_c = 667\:$Hz is shown in Fig.$\:$\ref{fig:fc_667Hz}. The process is driven by the pulses at $U=22\:$V and $f=500\:$kHz. The number of pulses in the series is  $N=500$; the active and passive times are $t_a =0.5\:$ms and $t_p=1\:$ms, respectively. The maximum stroke is $d_0=2\:\mu$m and in $1\:$ms the membrane returns to its initial state. The gas in the chamber is terminated much faster than in the case $f_c = 10\:$Hz. There is again a perfect repeatability of the membrane deflection for different cycles. If one would use this operation mode to drive a micropump, the pumping rate would be $8\:\mu$l/min with a dosage of $0.2\:$nl/cycle.

Video S4 shows a side view of the membrane for operation at $f_c=376\:$Hz. The actuator is driven by a series of $N=600$ pulses with the amplitude $U=21\:$V and frequency $f=500\:$kHz. The video was recorded at $6\:000\:$fps and the motion is slowed down 100 times. From this video one can extract independently on the interferometric measurements the deflection of the membrane. It is estimated as $d_0 = 2.7\pm0.7\:\mu$m. There is no one-to-one correspondence with the case shown in Fig.$\:$\ref{fig:fc_667Hz} since different sample and different operating frequency are used but the agreement is reasonable.

\subsection{Load, stability, and power consumption}

\begin{figure}[tbhp]
\centering
\includegraphics[width=.99\linewidth]{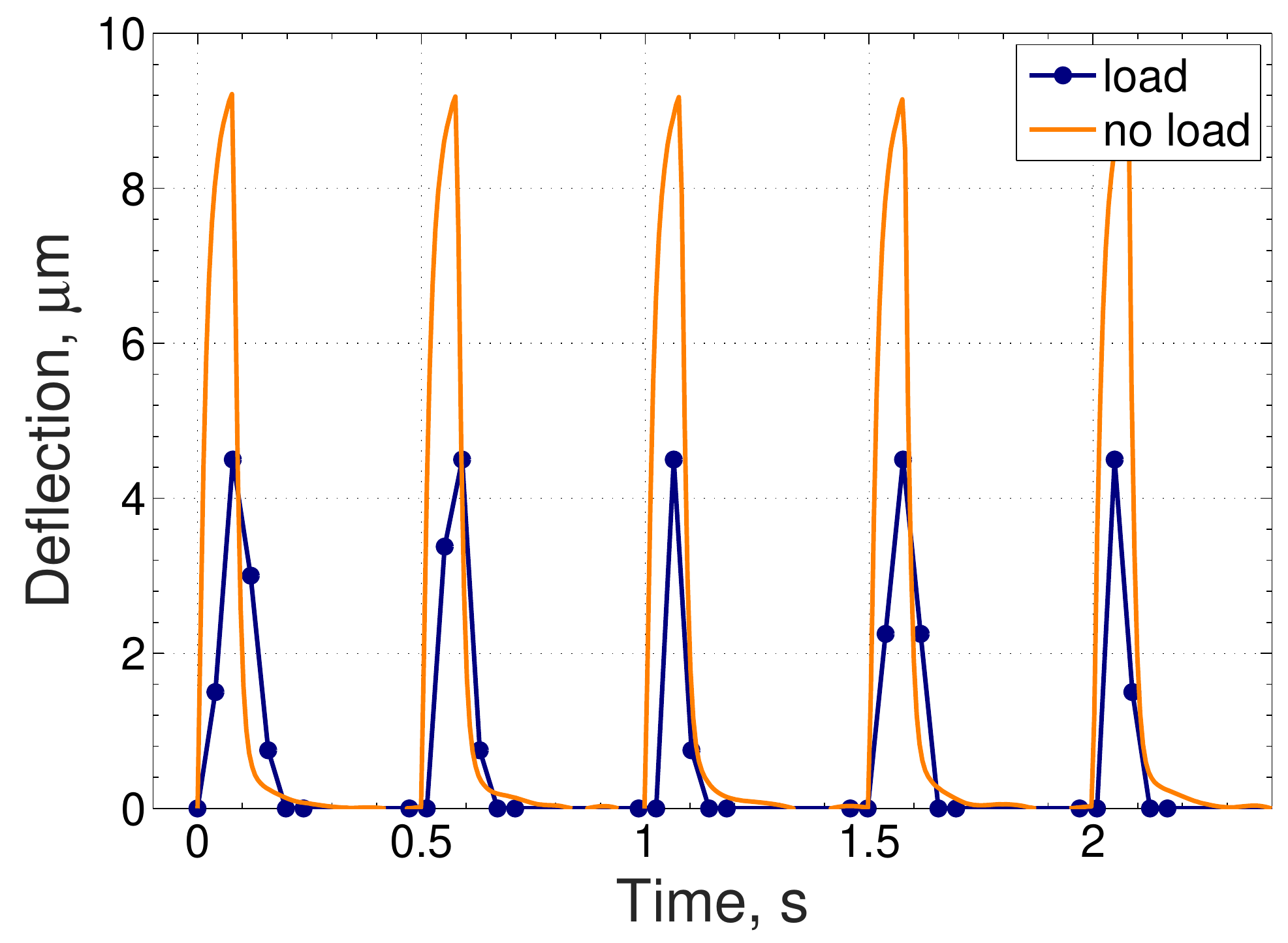}%
\caption{Operation of the actuator at $f_c=2\:$Hz loaded by the weight of the mirror (blue circles) and without the load (light brown line). }
\label{fig:load}
\end{figure}

In real application the actuator will produce mechanical work against a load force. To check the operation in load conditions we put the Si mirror on the membrane as shown in Fig.$\:$\ref{fig:load_scheme}. The load force on the membrane $F$ is produced by half of the mirror weight. The displacement of the mirror is measured interferometrically. The interferometer with quadrature signals cannot be used for this purpose because the sample in this instrument has to be mounted vertically. Instead we have used a simple one-signal interferometer and recorded the fringes pattern on video at $60\:$fps. Due to a relatively slow frame rate long series containing $N=80\:000$ pulses were applied at an operating frequency of $2\:$Hz. The process was driven at $U=14\:$V and $f=500\:$kHz. From the video it was possible to extract only a few deflection points per cycle. The same sample at the same running parameters was analysed without the load using the quadrature interferometer. The results are presented in Fig.$\:$\ref{fig:load}.

One can see that the load reduces the deflection of the membrane by half. It is possible to estimate roughly the extra pressure generated by the actuator to balance the mirror weight. This pressure is $P_b=F/\pi a^2$, where $F=0.6\:\mu$N and $a$ is the radius of the contact area between the mirror and the membrane. Assuming Hertzian contact \cite{LL7} the radius of the contact area is estimated as
\begin{equation}\label{contact}
  a=\left[\frac{3}{4}(1-\nu^2){\cal{R}}\frac{F}{E}\right]^{1/3},
\end{equation}
where $\nu$ is the Poisson ratio and $E$ is the Young modulus for PDMS and $\cal{R}$ is the curvature radius of the membrane. The latter can be calculated as ${\cal{R}}=(R^2+d^2)/2d$, where $R$ is the chamber radius and $d$ is the total deflection of the membrane counted from the neutral state. In our measurements we always keep some overpressure (by syringe) that the membrane is deflected without applying voltage. Typical deflection is $d=40\:\mu$m (as in frame 1, Fig.$\:$\ref{fig:explosion}(c)) and we find ${\cal{R}}=0.8\:$mm. For PDMS $\nu=0.5$; the Young modulus depends on the thickness of the membrane and for $30\:\mu$m it is about $E=1.4\:$MPa \cite{Liu2009}. For these values one finds $a=5.8\:\mu$m and $P_b=6\:$kPa. It means that the actuator working against the back pressure of about $6\:$kPa will reduce the stroke twice.

Stability of metallic electrodes is an important problem for durability of the devices. Most of the metals cannot withstand a high current density of the alternating polarity electrolysis. Titanium electrodes demonstrated the best results. In this work we compare durability of the actuators with Ti electrodes of different geometrical pattern: circular electrodes shown in Fig.$\:$\ref{fig:principle} and interdigital electrodes used previously for the micropump \cite{Uvarov2017}.  All tested samples operated $10-15\:$min in each regime and some of them run for about 1 hour of continuous operation. The interdigital electrodes demonstrate signatures of degradation at the end of the fingers already after a thousand cycles. The circular electrodes do not show signatures of degradation even after a million cycles. The reason for this difference is that in the case of circular electrodes the current is distributed more homogeneously than in the case of the interdigital electrodes. In the latter case the current density is the highest at the end of the fingers where the degradation proceeds.

An additional parameter that is important for practical realization of the actuator is the power consumption. Efficiency of the actuator discussed in this paper is rather low. Three major factors reduce the efficiency. First, one has to spent energy to decompose water into hydrogen and oxygen; second, a significant part of the generated gas disappears in the reverse reaction without producing mechanical work (nanobubbles containing stoichiometric mixture of gases); third, titanium oxide that is formed on the electrodes increases the operating voltage. All three factors enlarge the power consumption. This power can be calculated as $P_W = E_p(N/2)f_c$, where $E_p$ is the energy consumed per an active period. The energy is given by the integral $E_p = \int_0^T U(t)I(t)dt$, where $T=1/f$ is the driving period. The voltage $U(t)$ is presented by two rectangular pulses of positive and negative polarity; an example of the current $I(t)$ through the electrolyte is shown as inset in Fig.$\:$\ref{fig:fc_667Hz}. For the operating frequency $f_c=667\:$Hz we find $P_W = 300$mW. The actuation shown in Fig.$\:$\ref{fig:fc_10Hz} at $f_c=10\:$Hz gives $P_W = 79\:$mW.

The shape of the current influences the second factor and can be engineered to some degree. Narrow current pulses will reduce formation of nanobubbles containing both gases. Oxidation of electrodes also can be controlled in some way: we observed samples with operating voltage as low as $7\:$V (for long series of pulses) but most of the samples operate at $10-14\:$V. We expect that there is a room to reduce the power consumption for at least a few times but it needs further investigation.

\subsection{Discussion of the response time}

The most striking feature of the present actuator is a short response time. Moreover, comparison of operation at $f_c=10\:$Hz and $f_c=667\:$Hz shows that this time depends on $N$: the shorter the series of pulses the shorter the response time. For normal DC electrolysis hydrogen and oxygen microbubbles are formed. These bubbles cannot be easily terminated since the rate of the combustion reaction between the gases is negligibly low at room temperature and normal pressure. This is the main reason for a long response time observed in the DC electrolysis \cite{Neagu1996,Cameron2002,Hua2002,Ateya2004,Meng2008,Kjeang2009,Li2010,Yi2015}. In the case of the alternating polarity process only nanobubbles are formed. For these bubbles the surface-to-volume ratio becomes extremely high $S/V\sim 10^7\:$m$^{-1}$ and new surface assisted channels can be opened for the combustion reaction \cite{Prokaznikov2017}. This is the reason for a short response time in the alternating polarity process.

Since hydrogen and oxygen gases are not separated spatially, the nanobubbles containing both gases can be formed. In such a bubble the reaction is ignited spontaneously and the bubble disappears in tens of nanoseconds \cite{Prokaznikov2017} (for stoichiometric composition of gases) or ends up as a pure H$_2$ or pure O$_2$ bubble. The thermal effect of the reaction was observed  \cite{Svetovoy2011,Svetovoy2014} and investigated in more detail \cite{Jain2016}. The bubbles containing only hydrogen or only oxygen are collected in the system and they are responsible for the pressure increase and finally for the membrane deflection. While the electrical pulses are switched on the gas is produced by the electrodes and consumed in the reaction that can happen when two nanobubbles with different gases are met. In the steady state both of the processes equilibrate each other. When the pulses are switched off the gas termination dominates. For a long series of pulses the cloud of nanobubbles \cite{Postnikov2017} above the electrodes is larger and the density of nanobubbles is smaller than for a short series of pulses. The density is smaller because for long series the voltage amplitude is smaller and the current through the electrolyte is less than for short series producing similar deflection. This is the reason why the gas termination time is shorter for smaller number of pulses.

\section{Conclusions}

In this paper we demonstrated the microactuator that is using the alternating polarity electrochemical water decomposition as the driving principle. The main advantage of this process over the DC electrolysis is a very short time needed to terminate the gases produced by the electrolysis. In cyclic regime our actuator can work orders of magnitude faster than actuators using DC electrolysis. The shortest gas termination time reached so far is $1\:$ms. The cycles repeat themselves very precisely providing well controlled volume strokes in the range of picoliters.

In actuation by single series of pulses we demonstrated controlled explosions in the chamber between H$_2$ and O$_2$ gases and investigated the range of the parameters needed to produce the explosions. In the case of explosion the membrane jumps up to $90\:\mu$m for the time less than $100\:\mu$s. The cyclic actuation in the explosion regime is very attractive but more sophisticated electronics has to be used to control the process.

Most of the electrode materials are destroyed in the alternating polarity process by a high current density. Concentric titanium electrodes used together with sodium sulfate electrolyte demonstrated high durability due to relatively homogeneous current distribution.

\acknowledgments
VBS is grateful to Miko Elwenspoek who inspired this work. We thank H.S. Rho for help with the fabrication of PDMS membranes and A. Kupriyanov for assistance with the software. The work is supported by the Russian Science Foundation (grant 15-19-20003). VBS acknowledges partial support from the Dutch Technology Foundation (grant 13595).

\end{document}